\documentclass[fleqn,10pt]{wlscirep}
\usepackage[utf8]{inputenc}
\usepackage[T1]{fontenc}
\usepackage[english]{babel}
\usepackage{siunitx}
\usepackage{listings} 
\usepackage{subcaption} 

\definecolor{MyPlum}{RGB}{142, 69, 133}
\definecolor{MyOliveGreen}{RGB}{107, 142, 35}
\usepackage{xargs}
\usepackage[colorinlistoftodos, prependcaption, textsize=tiny, textwidth=1.5cm,color=red!20]{todonotes}
\newcommandx{\todoj}[2][1=]{\todo[author=Jero,linecolor=MyOliveGreen,backgroundcolor=MyOliveGreen!25,bordercolor=MyOliveGreen,#1]{#2}}
\newcommandx{\todof}[2][1=]{\todo[author=Jero,linecolor=MyPlum,backgroundcolor=MyPlum!25,bordercolor=MyPlum,#1]{#2}}
\newcommandx{\todol}[2][1=]{\todo[author=Lucas,linecolor=blue,backgroundcolor=blue!25,bordercolor=blue,#1]{#2}}
\newcommandx{\repl}[2][1=]{\todo[noline,backgroundcolor=blue!22,bordercolor=blue,#1]{#2}}

\title{Assessing the distribution of cancer stem cells in tumorspheres}

\author[1]{Jer\'onimo Fotin\'os}
\author[2]{Mar\'ia Paula Marks}
\author[1,*,+]{Lucas Barberis}
\author[2,*,+]{Luciano Vell\'on}
\affil[1]{IFEG-CONICET and FAMAF, Universidad Nacional de C\'{o}rdoba, C\'{o}rdoba, Argentina}
\affil[2]{Stem Cells Lab, IBYME-CONICET, Buenos  Aires, Argentina}

\affil[*]{lbarberis@unc.edu.ar, lvellon73@gmail.com}

\affil[+]{these authors contributed equally to this work}


\begin{abstract}
In previous theoretical research, we inferred that cancer stem cells (CSCs), the cells that presumably drive tumor growth and resistance to conventional cancer treatments, are not uniformly distributed in the bulk of a tumorsphere. To confirm this theoretical prediction, we cultivated tumorspheres enriched in CSCs, and performed immunofluorecent detection of the stemness marker SOX2 using a confocal microscope. 

In this article, we present a method developed to process the images that reconstruct the amount and location of the CSCs in the tumorspheres. Its advantage is the use of a statistical criterion to classify the cells in stem and differentiated instead of setting an arbitrary threshold. 
 
From the analysis of the results of the methods using graph theory and computational modeling, we concluded that the distribution of Cancer Stem Cells in an experimental tumorsphere is non-homogeneous. 

This method is independent of the tumorsphere assay being useful for analyzing images in which several different kinds of cells are stained with different markers.

\end{abstract}
\begin{document}

\flushbottom
\maketitle

\thispagestyle{empty}

\section*{Introduction} \label{s: intro}

Cancer Stem Cells (CSCs) are defined by their capability to \emph{replicate} making exact copies of themselves or to \emph{differentiate} giving rise to phenotypically diverse cells. They are resistant to conventional anti-cancer treatments, being thus implicated in disease recurrence and metastasis\cite{Al-Hajj, Kakarala}. One of the main barriers to the development of CSCs-targeted therapies is their scarcity in vivo, limiting their availability as experimental systems for pharmaceutical development, and raising the need for production at a scale large enough to fulfill academic and industry requirements. A common solution is the anchorage-independent growth of cancer cells to generate a 3D culture of epithelial cells\cite{Dontu} enriched in cells with CSCs-like properties since most epithelial cells are killed by anoikis\cite{Ehmsen}. Thus, taking into account the limitations of this assay\cite{Pastrana}, the resulting tumorspheres are formed by the clonal expansion of a single cell, instead of the self-aggregation of existing cells.

Still, routinely used cell culture techniques are material- and labor-consuming tasks that generate a great amount of inter-culture variability and contamination risks. Moreover, traditional cell culture at a large scale is also cost-ineffective in terms of the high investment in cell culture media and growth factors. 
In this context, developing forefront, high-throughput screening platforms to identify cytotoxic inhibitors and/or differentiation-promoting agents targeting CSCs becomes of paramount importance. This would require the optimization of a series of bioprocesses that enable the massive culture of undifferentiated cancer cells and the analysis of high volumes of data. 

One critical downstream part of such bioprocesses is to assess the cellular response in terms of viability and/or stemness markers, which requires external software for image analysis and segmentation to quantify the relative differences among treatment groups. Even though high-throughput cytometric methods have been developed\cite{Kessel}, there is still the need to know the exact identification, location, and targeting of putative CSCs. This would help the modeling of CSC dynamics, allowing the development of cost-effective and predictive tools for the examination of tumor evolution and response to therapy.

Mathematical modeling has contributed successfully to the study of tumor growth and treatment response. One shortcoming is the risk of parameterization from multiple sources, thus mixing cancer types, experimental conditions, and even spatiotemporal scales \cite{Brady2019}. Here, we attempt to overcome this issue by establishing a method of image processing that fits biological data with previously developed mathematical models for the distribution of CSCs in tumorspheres \cite{barberis2021percolation}. Taking this into consideration, we chose tumorspheres as a simplified 3D tumor model since they are cellular structures that are generated from a variety of tumors from epithelial tissues, such as breast, lung, prostate, or colorectal cancer \cite{Zanoni2020}. Even though the optimal conditions for culturing tumorspheres may differ among tumor types, this experimental system can recapitulate the physicochemical gradients from the spheroid periphery to its core and mimic, to some extent, mechanical properties and cell-cell interactions of avascular tumor mass microregions \cite{Rolver2019,Zanoni2020}. 

In our case, mathematical methods allowed us to estimate the expected fraction of CSCs for tumorspheres in different culture conditions\cite{benitez2019, barberis2021diff, benitez2021} and the effect of specific therapies on their development\cite{fotinos23} among other theoretical results\cite{Condat2006, Delsanto2008, Menchon2011, Barberis2015}.  

We have also computationally simulated the growth of a colony of cells in two dimensions using an Agent-Based Model (ABM) that mimics basic features of CSCs proliferation to form a spheroid\cite{barberis2021percolation}. The simulated spheroid grows from a single CSC and the cells can undergo mitosis at a fixed rate (the population doubling time or PDT). Depending on the intrinsic and extrinsic (micro-environment) signals, the CSC will replicate generating another CSC with a certain probability $p_s$, yielding a \emph{differentiated cancer cell} (DCC) otherwise. This simplified model allowed us to estimate the total number of CSCs, the fraction of CSCs situated on the periphery of the colony, and the size of the whole spheroid, showing that these traits are dependent on the replication probability ($p_s$) of the CSCs. Indeed, simulating with intermediate replication probabilities, we observed active CSCs at the border of the colony and detected that they form a path that links the center of the colony with its border. Furthermore, an increase in the replication probability led, as expected, to a large CSCs population that overtook the system. This last situation may describe most experimental conditions used for culturing tumorspheres \cite{chen2016,wang2016, Dontu, Leis, Marks2024} and agrees with our previous mathematical models \cite{benitez2019, barberis2021diff, benitez2021}.

Inspired by these simulations, we generated tumorspheres from MCF-7 cells and analyzed them by confocal microscopy.  We developed an advanced computational method to detect the expression and distribution of the stem-like cells, SOX2-positive cells, under the assumption\cite{Leis} that this stemness factor is expressed in tumorspheres from cell lines and primary cultures. The processed images were used to study the distribution patterns of CSCs employing statistical tools that validate our main computational finding: that CSCs are heterogeneously distributed in a tumorsphere. 

Summarizing, we measured a non-homogeneous distribution of CSCs in tumorspheres after a computational model predicted it. In particular, we report a protocol to analyze confocal images of tumorspheres that were stained with a stemness marker, to statistically infer the more suitable threshold to determine which cells in the culture belong to the stem phenotype. We highlight the capability of our method to extract data from the experiments allowing direct comparison with simulated cultures. With this purpose, theoretical work was useful to drive experiments and, conversely, ad hoc experiments became essential to develop more accurate theoretical work.

\section*{Results} \label{s: results}
According to simulations\cite{barberis2021percolation}, CSCs must be connected in tumorspheres forming ``paths'' that join the center of the spheroid with its border. To assess this, we performed tumorspheres assays where MCF-7 cells in suspension cultures proliferate in a solution enriched in growth factors to ensure a large fraction of CSCs. After 9 days of growth, we collected the spheroids and attached them to a slide by cytocentrifugation. Due to the centrifugal forces, their spheroidal shape may be lost, becoming discs smashed against the slide, however, the structure of the tumorsphere is maintained. We stained the slides to detect the location of all cellular nuclei (DAPI), and the corresponding primary and secondary antibodies needed to detect the position of the stem cells (SOX2). Finally, we took pictures on several focal planes (\emph{slices}) of the spheroid obtaining a full 3D reconstruction. Because of the smashing of the spheroids, only one or two slices of the image were needed to observe all the cells present in them. These images were processed with a Data Science approach, using computer vision and statistical analysis techniques, among others. The result of the process allows us to mathematically reconstruct the discs, specifying the position of all their cells, marking those that are candidates to be CSCs and statistically deciding which ones are truly CSCs.
The experiment and the image processing procedure are fully detailed in the Methods section.

\subsection*{The Cancer Stem Cells distribution. }

\begin{figure}[!ht]
    \includegraphics[width=\linewidth]{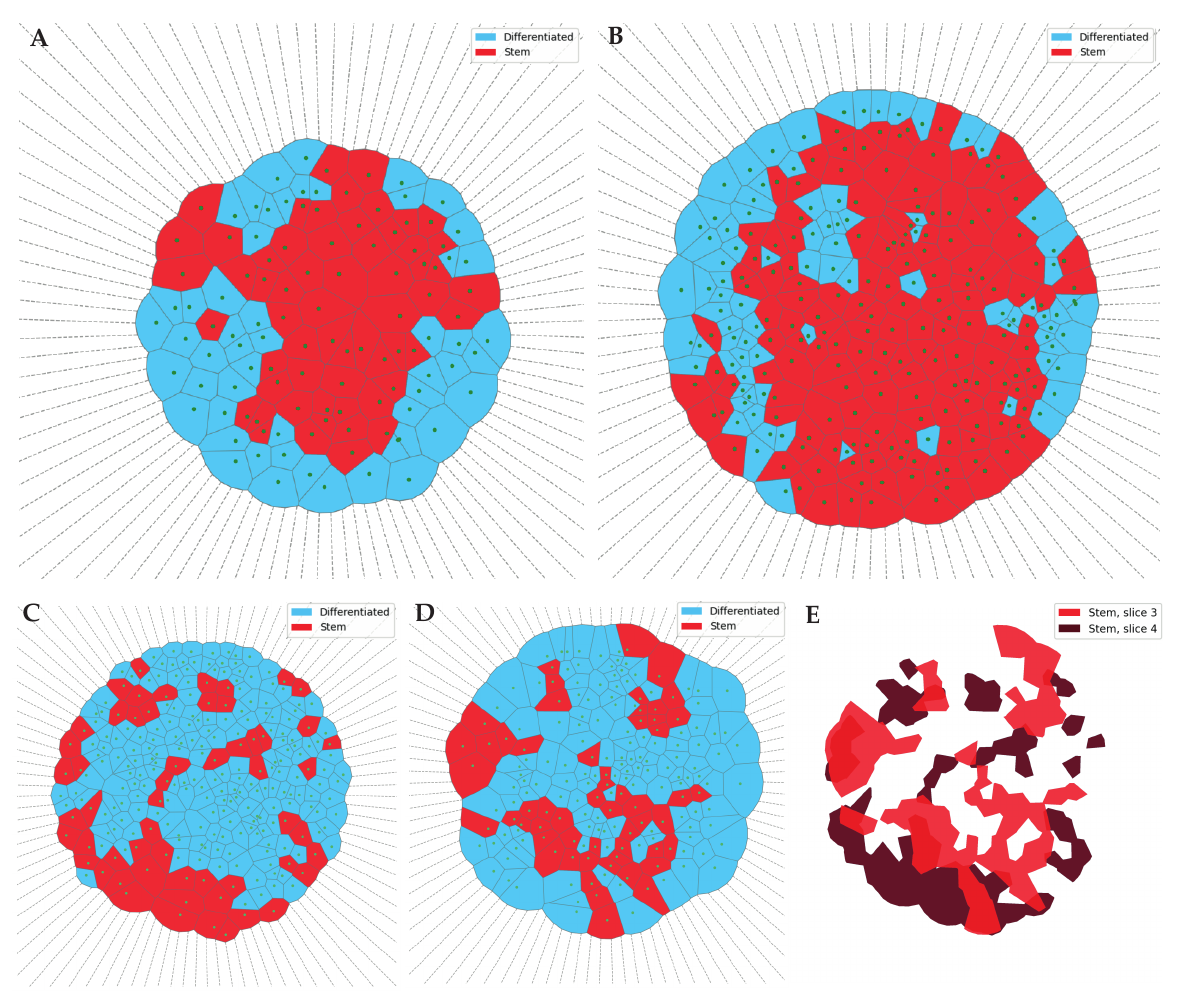}
    \caption{\emph{Reconstruction of the spheroids' slices after image processing.} For different slices (i.e., focal planes) of the experimental spheroids, the cell's area is approximated by polygons. The polygons representing CSCs are colored in red, and the ones representing DCCs, in blue. Note that the CSCs are not uniformly distributed in any image, and form patches at the border of the aggregate. The panels correspond to: \textbf{A} \textsf{ Sph4, slice 2}; \textbf{B} \textsf{ Sph3, slice 3}; \textbf{C} \textsf{ Sph1, slice 3}; \textbf{D} \textsf{ Sph1, slice 4}; \textbf{E} Overlay of CSCs of slices 3 and 4 of \textsf{ Sph1}. This last panel shows how stem cell clusters are connected through the slices. }\label{fig: blue red}
\end{figure}

Our main results are summarized in Fig. \ref{fig: blue red}, in which we can observe the distribution of the two cell phenotypes:  CSCs in red and DCCs in blue. We present three representative examples belonging to the reconstruction of three different spheroids. After the filtering and reconstruction process of the confocal images, we obtained pictures of the cell's distribution as a Voronoi tessellation (see Methods section). That is, each cell is represented, in Fig. \ref{fig: blue red}, as a polygon whose centroid coincides with the centroid of the corresponding cell in the pictures. Note that this representation is just an approximation of the shape of the cell that allowed us to quantify the SOX2 content inside the cell and decide, using statistical tools, if the cell displayed a stem or a differentiated phenotype. After the whole process, the CSCs become represented by red polygons and the DCCs by blue polygons. 

\subsubsection*{Patches in the border}

Our first example, labeled \textsf{ Sph4, slice 2} as in raw data files, a spheroid that has not grown much is depicted in Fig. \ref{fig: blue red}A. Considering the different proliferative capacities of the cells in these culture conditions and the possibility of differential access to the growth factors, we expect inter-spheroid variability in both the number of cells and the CSC fraction. In this case, the spheroid had less than a hundred cells, with half of them being CSCs, cf. Table \ref{tab: experimental values}. There is an evident higher concentration of the CSCs in the core of the spheroids, which could be explained by the model used in our simulations: the first CSC has a given chance of generating another CSC, but when one of these CSCs differentiates the first time, its lineage will only contain DCCs that will surround its stem mother. The only exception is when a DCC cannot undergo mitosis in a nearby place, already occupied by a CSC. Indeed, this CSC will continue generating other CSCs until one of them becomes differentiated, leaving in the process a path of CSCs among DCCs. This is exactly what is shown in panel A of Fig. \ref{fig: blue red}: at an early stage, red CSCs become surrounded by blue DCCs in the lower portion of the spheroid. However, three CSCs were able to initiate paths that extend to the periphery of the spheroid. The result is that CSC will form patches in the border of the spheroid.

A larger growth rate is associated with the case labeled \textsf{ Sph3, slice 3} shown in Fig. \ref{fig: blue red}B, where the spheroid has more cells than the one shown in panel A. Its CSC fraction is 64\%, larger than the 54\% of the previous case, a result almost trivial that agrees with our most used modeling hypotheses: the growing rate can only be measured for the bulk of the cells\cite{benitez2021}, thus, the outcome of the mitosis of a CSC is given just by a probability rather than a specific growth rate for each cell phenotype. Moreover, while we expect the quantitative results reported in Barberis, 2021\cite{barberis2021percolation} to be significantly different for three dimensions, the qualitative ones are likely to remain the same. The CSCs would form paths and be heterogeneously distributed, forming ``patches'' at the border (surface) of the spheroids after approximately one week.  

In this example, we can also appreciate the deformation produced by the smashing against the slide. Some DCCs appear in the bulk of the disk surrounded by CSCs. These DCCs could have initially belonged to the border of the spheroid and reached the center after centrifugation. In this manner, it is easy to understand why cells closer to the center of the disk could be a mix of original center cells and some of the periphery. However, there is no way (it is very unlikely that cells at the border?-no me gusta usar way en los papers, me parece muy street language) in which cells at the border of the spheroid could originally be from near the spheroid center.

\subsubsection*{Two layers sample}

As mentioned, our method for placing the sample on the slide alters the shape of the spheroids, projecting the cells onto a disc. Depending on the forces acting on the spheroids, sometimes their cells can not push all their neighbors apart, piling up on top of them. As a consequence, we need to use the images of more than one slice to recover all the cells that constitute the spheroid. 

An example of this is given by \textsf{ Sph1} which had two layers of cells superimposed and required two slices of confocal imaging to access all the cells. In panels C and D of Fig. \ref{fig: blue red} we depict the obtained cell distribution for the upper and bottom layers respectivelly. As observed by direct inspection, and reported in Table \ref{tab: experimental values}, this spheroid has a larger amount of cells than those in the previous examples, but a much smaller CSC fraction. Thus, even though CSCs are thought to drive tumor growth, their abundance may not be directly proportional to it. This can be understood through our modeling hypothesis about CSC reproduction. When CSCs have a small chance to replicate  becoming differentiated at early times, they are likely to become quickly surrounded by DCCs. This is again consistent with our description of the replication rate as a probability that models the chances of a CSC becoming undifferentiated either because of extrinsic or intrinsic causes. 

In the two-layer disc, we have an example of such a situation, where it becomes difficult to follow the CSCs' path. The CSCs seem to be less connected and not as clustered as in the previous cases, and might even seem more randomly distributed. Nevertheless, this case presents the opportunity for a better reconstruction of the spheroid due to the smaller loss of spatial information.  Indeed, if we subtract the DCCs from the snapshots in panels \ref{fig: blue red}C and \ref{fig: blue red}D, and overlay the remaining CSCs, we obtain the distribution shown in panel \ref{fig: blue red}E. Note that we colored the bottom layer cells (against the plate) in dark red, and the upper layer in light red. The result is now evident, most of the CSCs are indeed connected, forming a path along the two layers as we expected from the simulations.

\begin{table}[htbp]
\centering
\begin{tabular}{|l|c||c||c|c|c|}
\hline
\textbf{} & \textsf{ Sph4, slice 2}& \textsf{ Sph3, slice 3} &  \textsf{ Sph1, slice 3} & \textsf{ Sph1, slice 4} &\textsf{ Sph1, both slices}\\
\hline
\hline
Radius (\textmu m)   & 87.58& 102.04 & 113.25 & 99.07 &---\\
\hline
Total Cells & 112  & 240 & 183 & 252&  435\\
\hline
Stem Cells   & 59 & 155& 55 & 56& 111\\
\hline
Differentiated Cells  & 53& 85 & 128 & 196& 324 \\
\hline
Stem Cell Fraction  & 53\%& 64\% & 30\% & 22\% & 26\% \\
\hline
\hline
\end{tabular}
\caption{\emph{Experimental size and composition of the spheroids, recovered after processing the images.} For each slice, we report the radius of the spheroid, the total number of cells (as well as the number for each population), and the fraction of them that are stem cells. In the case of  \textsf{ Sph1}, we reported the measured values for each slice and the total values, summing up the data of the two slices in the last column. Note that this spheroid, being the larger one, has the smaller CSC fraction, in agreement with the simulations. }
\label{tab: experimental values}
\end{table}

\subsection*{Non-randomness of the CSC's distribution}

\begin{figure}[!ht]
        \centering
        \includegraphics[width=\textwidth]{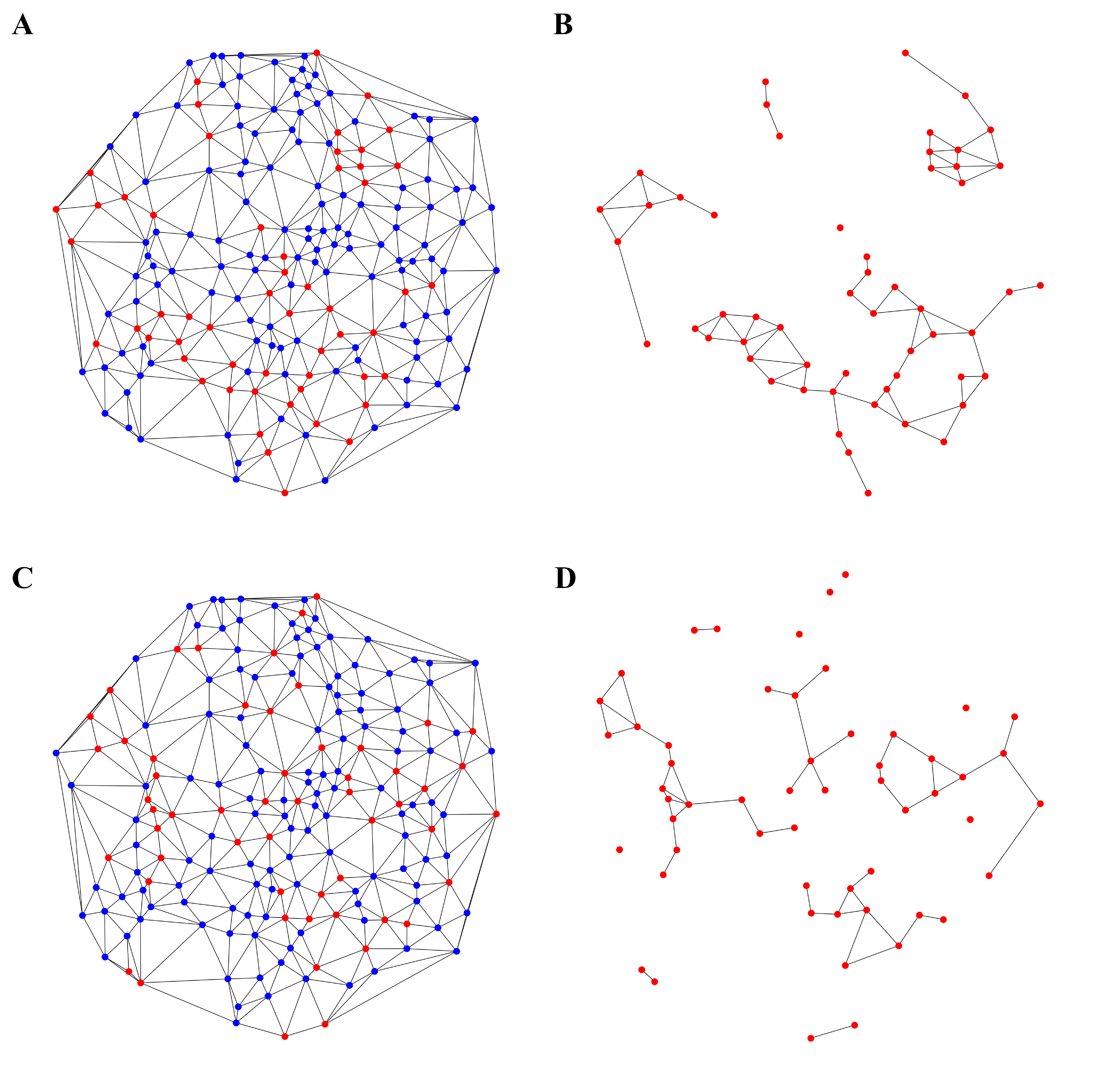}
    \caption{\emph{Comparison of an experimental network and its stem subgraph, with a copy of it with randomly located CSCs.} {\bf A} The reconstructed network for \textsf{ Sph1, slice 3} with its CSCs plotted as red dots, and its DCCs plotted as blue dots. Neighbor cells are connected with gray links. {\bf B} The CSCs subgraph corresponding to the network in A is obtained by deletion of the DCCs (blue) nodes. The number of stem/connected components is 5. {\bf C} The random graphs are constructed by randomly redistributing the red and blue dots on the same network. {\bf D} The  CSCs subgraph corresponding to the randomized network in C has a different number of connected components, 13 in this case. In most random cases, this number is larger than in the experimental network. }
    \label{fig: Delaunay sph1 slice3}
\end{figure}

We measured some extra features that allow us to be sure that our findings have statistical significance, even for just a sample. The tessellation generated by this method defines the location of a cell and its neighbors, allowing to make a graph of the connections between cells. An example of this is depicted in Fig. \ref{fig: Delaunay sph1 slice3}A for \textsf{ Sph1, slice 3}, where the CSCs are represented by red dots, and DCCs by blue dots. If two cells are neighbors, they are linked with a black line. With this graph, we can use network theory to measure how connected are the CSCs inside the spheroid structure. 

At first glance, is easy to see that CSCs form the expected path. Moreover, we assessed if this is true by measuring how many CSCs paths are present in the graph and, most importantly, how apart one is from the other. The physics of complex systems has developed tools to analyze graphs and collect nontrivial information. In our case, we determine if the paths formed by CSCs could appear by chance, or if there must be a defined mechanism behind them. As explained in the Methods section, we test for homophily in the reconstructed networks for the experimental systems. That is, the tendency of cells to neighbor other cells of the same phenotype. To characterize this, we use the \emph{assortativity coefficient}, which is greater than zero for networks that present homophily, and results positive for our experimental cases. Also, the \emph{homophily ratio}, the fraction of pairs of neighbors of the same kind, gives relatively large values (meaningful comparison to be established later). Both results mean that the nodes prefer to be attached to nodes of the same kind. The measured values of these quantities for the four examples shown here, are reported in Table \ref{tab: ensemble statistics} under the columns labeled `experiment'.   

For a meaningful comparison and interpretation of these values, we used a classical approach from the physics of complex systems. We maintain the connections between the nodes of the previously obtained networks, but this time we arbitrarily set which ones are the CSCs, keeping their total number equal to the corresponding experimental network, cf. Fig. \ref{fig: Delaunay sph1 slice3}C. These random distribution of the CSCs is one of the most used hypotheses in mathematical modeling. This will also resemble the case of CSCs that can travel inside the spheroid. We obtained 10000 of these networks with randomly distributed CSCs, and we calculated the quantities mentioned before for characterizing their homophily. Then, we performed statistical tests (always using significance $\alpha_t=0.001$) to compare these values with the ones obtained for the experimental cases. These tests tell us whether the quantities measured in the experimental networks significantly differ from the same quantities measured in the random networks. In Table \ref{tab: ensemble statistics} we reported the averaged results over all the realizations including their standard deviation. The assortativity coefficient is now zero meaning that, as expected, there is no preferential attachment of cells of the same kind when we randomly locate the CSCs. The random homophily ratio is a bit larger in the two-slice spheroid but still lower than the experiential case. Thus, experimental values tell us that daughter cells indeed stay close to their parent cell. 

A way to double-check this result is by looking at the stem subgraphs. As shown in Fig. \ref{fig: Delaunay sph1 slice3}B, this is the network formed just by the CSCs and the connections between them. We measured the degree distribution of these subgraphs, deriving information such as the average number of connections that the nodes have. We also calculated the number of connected components of this graph, which is the number of separated groups of CSCs. The more clustered the CSCs, the less the connected components. These values are also reported in Table \ref{tab: ensemble statistics} both for the experimental and the average over the random networks.

For \textsf{ Sph3} and \textsf{ Sph4} the large CSC fraction made the experimental and ensemble values statistically indistinguishable from each other. This means that, because we have more than half of CSC in the spheroid, there is no chance to get many isolated CSC, independently of the way they are distributed. 
On the other hand, for \textsf{ Sph1} the measured number of connected components was significantly smaller than the one measured for the random ensemble, stressing that despite being fewer CSCs than DCCs, they are indeed connected. A close inspection of Figs. \ref{fig: Delaunay sph1 slice3} and \ref{fig: blue red} reveal that most of the connected components are separated from each other by just one DCC as predicted by simulations. Thus, the number of \emph{effective} connected paths, in the sense defined in \cite{barberis2021percolation}, should be even smaller.

The mean degree of the experimental stem subgraphs was significantly higher than their corresponding random averages. This implies that CSCs are significantly more connected between them than expected if located randomly. This, in turn, reinforces our previous conclusion about the tendency of neighboring cells of the same phenotype. 

\begin{table}[htbp]
\centering
\begin{tabular}{|l|c|c|c|c|}
\hline
 & \multicolumn{2}{|c|}{\textsf{ Sph4, slice 2}}  &  \multicolumn{2}{|c|}{\textsf{ Sph3, slice 3}}\\
\hline
 &experimental &random& experimental & random\\
\hline
Assortativity Coefficient &
0.39 & -0.010$\pm$ 0.04& 0.47 & -0.01 $\pm$ 0.06 \\
\hline
Homophily Ratio &  0.74&0.50 $\pm$0.03&0.73&  0.54$\pm$  0.02  \\
\hline
Stem Connected Components &2& 3.7 $\pm$ 1.5 &2&  1.1$\pm$  1 \\
\hline
Degree of Stem Subgraph &4.5& 3.0$\pm$ 0.2  &4.8& 3.7$\pm$ 0.1\\
\hline
\hline
\hline
 & \multicolumn{2}{|c|}{\textsf{ Sph1, slice 3}} & \multicolumn{2}{|c|}{\textsf{ Sph1, slice 4}}  \\
\hline
 & experimental & random& experimental&random \\
\hline
Assortativity coefficient &0.22 & -0.01 $\pm$ 0.04   &0.26& -0.01$\pm$ 0.01  \\
\hline
Homophily ratio  &0.67& 0.58$\pm$ 0.02 &0.74& 0.65$\pm$  0.01 \\
\hline
Stem connected components &5& 17$\pm$ 3 &12& 25 $\pm$ 3.3 \\
\hline
Degree of stem subgraph &2.7& 1.7$\pm$ 0.2 & 2.64 & 1.3$\pm$ 0.2  \\
\hline

\end{tabular}
\caption{\emph{Homophily comparison of the experimental and the random ensemble of networks.} Network properties from the experimental networks are shown in columns labeled `experimental', and the ones for networks with randomly distributed CSCs, in columns labeled `random'. For these last ones, we report the mean values and their standard deviation. The assortativity coefficient is zero for the random cases, stressing that there is a no preferential attachment, contrary to what happens in the experimental cases where positive values are always found. For \textsf{ Sph4} and \textsf{ Sph3} the connected components are similar due to the large amount of CSCs. }
\label{tab: ensemble statistics}
\end{table}

\section*{Discussion}

As expected from our simulations, CSCs are not randomly distributed in the tumorsphere, they rather form a path and tend to interact with cells of the same type. This might reflect the phenomenon of stem cell competition. Briefly, stem cells reside within specific microenvironments (niches) which provide restricted maintenance signals and limited physical space, and consequently, stem cells are constantly competing with their neighbors for niche occupancy \cite{Stines2013}. In the non-neutral stem cell competition, a fraction of stem cells gain a fitness advantage or disadvantage over their neighboring stem cells. Thus, the more competitive fraction overtakes the niche, sometimes disrupting it and leading to diseases such as cancer \cite{Johnston2009}. Indeed, mutations that affect cell fitness either in development or homeostasis can lead to a competitive growth advantage and potentially clonal expansion. Alternatively, non-neutral stem cell competition eliminates ‘‘unfit’’ clones, for instance when aneuploid cells are depleted during development \cite{Derks2023}. However, in the case of tumorspheres, it is difficult to assume the type of stem cell competition due to the selection already imposed by the culture conditions. It is possible, however, that the hypoxic cores in the tumorsphere can trigger differential cell responses leading to drug resistance \cite{Fisher2020} and these CSCs, more competitive, deplete the drug-sensitive cell populations and eventually reach the surface of the spheroid through replication. One example of this has been reported in tumorspheres from the breast cancer cell line BT474, in which upregulation of HER2 expression led to a hypoxia-conditioned breast CSCs population with increased resistance to trastuzumab\cite{Rodriguez2018}. Even though these effects have been observed in large-sized, multicellular spheroids with significant oxygen, nutrients and metabolite gradients, the fact that small spheroids (25-50 cells) contain cells more resistant than monolayers to chemotherapeutic agents has been repeatedly observed long ago \cite{Olive1994}. This suggests that other factors, possibly mechanical/geometrical restraints imposed by the shape of the tumorsphere affect CSCs location and cell-cell interactions. Therefore, mathematical models of how different subpopulations of cells interact in the context of a multicellular aggregate of spheroidal shape and validating these models with biological data may contribute to developing cost-efficient and predicitve tools. In this regard, image processing is useful to obtain information that is highly relevant for setting up simulations, formulating mathematical models and fitting their results. Here, we measured both, the total amount of cells in each spheroid and the respective CSC fraction, information that will lead to improving the accuracy of mathematical models such as the ones summarized in \cite{barberis2021diff}. Furthermore, we estimate that the growth rate is around $r=$\SI{1.1}{cell/day}, which allows us to establish and set the temporal scale of simulations as those in \cite{barberis2021percolation}. Of note, this value is in agreement with the one obtained by fitting experimental data with mathematical modeling\cite{benitez2021} and encourages us to extend to three dimensions our mathematical and computational models.

\section*{Conclusion}

The method of image processing presented here is devoted to recognizing the location of the CSCs in microscopy images. Its originality lies in the fact that we are not just looking at where the SOX2 fluorescence is high enough under a subjective criterion. The key is the use of the Gaussian Mixture Model to fit the data and separate, with a statistical criterion, both cell populations. Having done this, the way of depicting the spheroid or what is done with the resulting data will depend on the questions that researchers have in mind. In our case, we presented an example that compares the experimental CSC distribution in tumorspheres with a previous computational model. Beyond the fact that we are pleased to find that \emph{in silico} and \emph{in vitro} experiments give similar outcomes, the method proposed here is straightforwardly applicable to other similar experiments. Indeed, to definitively assess the CSCs distribution in a spheroid, we must be able to put into the microscope non-deformed spheroids and take pictures of several slices of them.
Furthermore, according to our computational simulations, the confidence in our result will exponentially increase with the cultured time. Indeed, we are carrying out full 3D simulations of tumorsphere assays. The preliminary results still support the finding that CSCs are heterogeneously distributed inside the spheroid. However, the probability of finding CSCs on the border of the spheroid seems to be significantly enlarged. These simulations plus the help of confocal images of several slices of a non-deformed spheroid, will be the basis for studying further therapies focused on treating cancer stem cells.

\section*{Methods} \label{s: methods}

\subsection*{Cell Culture and tumorsphere assay}

Human breast cancer-derived cell line MCF-7 was maintained in DMEM/F12 complete medium (DMEM F12 + FBS 10 + 1 Glutamine) at $37 \ ^{\circ} C$ in a humified 5 CO2 atmosphere. The tumorsphere assay was performed according to modifications \cite{Leis} from the original protocol by Dontu \emph{et al.}\cite{Dontu}. Briefly, 6-well plates were treated with poly(2-hydroxyethyl methacrylate) to prevent cell adhesion. MCF-7 cells were seeded at 3000 cells/ml in a complete mammospheres medium (DMEM/F12 + B27 2 + glutamine 1 +20 ng/ml EGF + 20 ng/ml bFGF). Cells were incubated at $37\ ^{\circ} C$  in a humified 5 CO2 atmosphere for 7-9 days, adding 0.5 ml of medium with growth factors every 48 hs. The resulting mammospheres were collected and attached by cytocentrifugation onto slides for immunofluorescent detection of SOX2.

\subsection*{Immunofluorescence}

Immunofluorescent detection of SOX2 in mammospheres was performed as described in Leis \emph{et al.} \cite{Leis}: slides were fixed in methanol and then washed 3 times in washed solution (PBS-BSA 0.1) for 5 minutes and permeabilized with PBS-BSA 0.1 + 0.3 triton X100 for 30 min at RT. Following permeabilization, the slides were incubated in blocking solution (PBS-BSA 0.1 + Normal Goat Serum) and the corresponding primary antibody (anti-SOX2, cat \# PA1-16968, Thermo)  ON at 4C. Next, the slides were washed 3 times in washing solution for 5 minutes and incubated with the corresponding secondary antibody (anti-rabbit Alexa Fluor (TM) 555 cat \# A31572, Life Technologies).   

\subsection*{Confocal Imaging}

Images were acquired with a Zeiss LSM 880 Laser Scanning Confocal microscope with a 20X objective. The corresponding lasers and photomultipliers for excitation and detection of Alexa Fluor 555 and DAPI were used. Further specifications regarding the obtained images are available at the public repository specified in the Additional information section.

\subsection*{Image Analysis}
In the files obtained by digital imaging through confocal microscopy, we recognize the number, position, and extent of cells in the spheroids, and tag CSCs and DCCs according to the amount of SOX2 fluorescence using a statistical criterion.  In the following, we describe the whole process and show the partial results on spheroid number 1, \textsf{Sph1} hereafter, as an example. We began with the corresponding \textsf{.czi} file that contains data on three channels labeled \emph{optic}, \emph{DAPI} with the fluorescence of the nuclei, and \emph{SOX2} with the SOX2 fluorescence. The data are files containing the position of the pixels in three spatial dimensions $(x, y, z)$ with 2292 pixels representing $ 0.12 \mu \text{m} $ for each side in the horizontal $x-y$ plane, and a thickness of 9 pixels representing  $2 \mu \text{m.}$, in the $z$ direction. For each pixel, the corresponding value for each channel is included. We start analyzing each channel separately, and then we merge their information. 

\subsubsection*{The SOX2 Channel}
\begin{figure}[!ht]
    \centering
    \includegraphics[width=\linewidth]{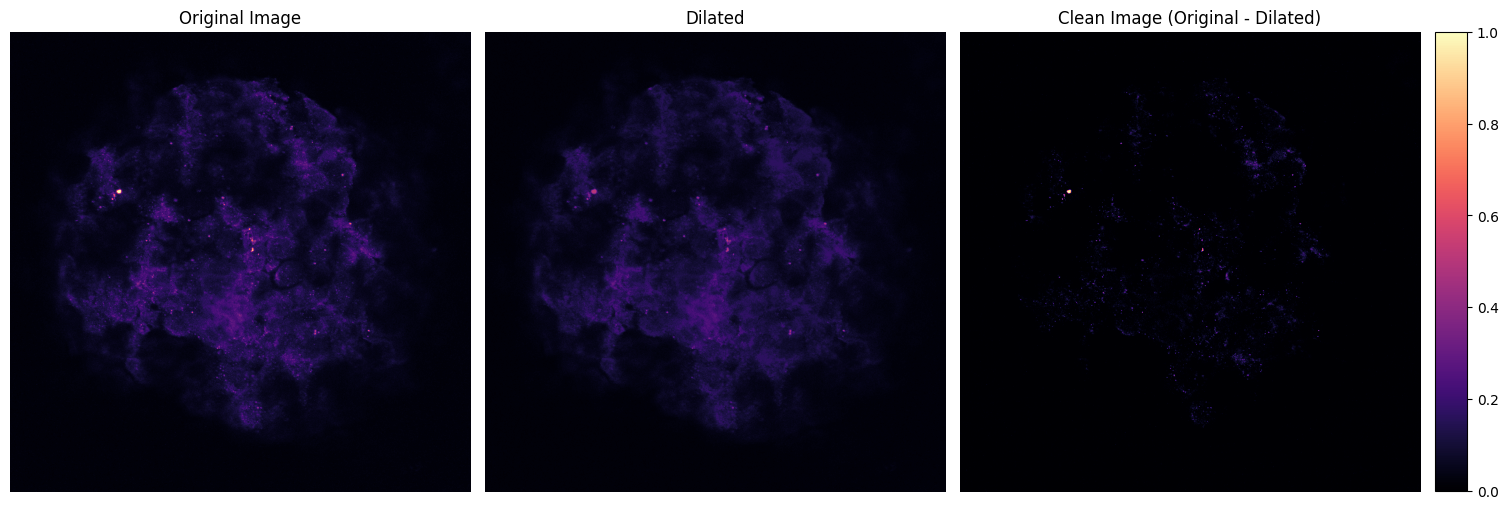}
    \caption{\emph{Cleaning of the SOX2 channel.} From left to right: the original image, a dilated version of it (a blurred version or background noise), and the cleaned image as the difference between them. The original image corresponds to the SOX2 channel of \textsf{Sph1, slice 3}. Note that the dilated image allows us to remove a large amount of non-informative pixels.  }
    \label{fig: sox2 before and after}
\end{figure}

The technique reveals the spots where the marker was attached to the cells, getting rid of the diffused SOX2 in the bulk. To achieve this, we perform a reconstruction by dilation, a morphological image processing technique that provides a denoised version of the image, by subtracting from it a blurred version of the original one. Fig. \ref{fig: sox2 before and after} shows the original image in the left panel, the dilated image at the center, and the resulting cleaned image in the right panel.
Specifically, we used a \lstinline{Scikit-image}'s implementation\cite{scikit-image} of the morphology algorithms. For further information, see \href{https://scikit-image.org/docs/stable/auto_examples/color_exposure/plot_regional_maxima.html#sphx-glr-auto-examples-color-exposure-plot-regional-maxima-py}{\lstinline{scikit-image}'s documentation}). 

\subsubsection*{The Nuclei Channel}
For this channel, we use the nuclei to individualize and identify the cells. Specifically, we performed instance segmentation on the nuclei, extracting geometrical features such as the area and the coordinates of the center of the segmented objects, according to the following steps: 

\begin{enumerate}
    \item Contrast Limited Adaptive Histogram Equalization (CLAHE): Used for local contrast enhancement (see \href{https://scikit-image.org/docs/stable/api/skimage.exposure.html#skimage.exposure.equalize_adapthist}{documentation}) makes the edge recognition easier for the segmentation algorithm.
    
    \item Morphological processing: To further improve the edge recognition, we performed a morphological opening (erosion followed by dilation), followed by an area closing (similar to a morphological closing, dilation followed by erosion, but using a deformable rather than a fixed footprint). The opening helps separate objects that may be in contact. The area closing removes small dark structures to avoid single nuclei being segmented as many objects due to dark spots within them (see \href{https://scikit-image.org/docs/stable/api/skimage.morphology.html#skimage.morphology.area_closing}{documentation}).
    
    \item Bilateral Denoising: An edge-preserving bilateral filter denoises the image, averaging pixels based on their spatial closeness and radiometric similarity. This filter works both to reduce the noise introduced by the morphology operations and by out-of-focus objects (see \href{https://scikit-image.org/docs/stable/api/skimage.restoration.html#skimage.restoration.denoise_bilateral}{ documentation}).
    
    \item Instance Segmentation: For identifying the nuclei, we used the \textsf{2D\_versatile\_fluo} model (see \href{https://github.com/stardist/stardist}{documentation}). This is a \textsf{Stardist} \cite{schmidt2018} convolutional neural network with a U-Net architecture, trained on fluorescence microscopy images similar to the ones of our experiment. 
    
    We chose this model instead of instance segmentation with bounding boxes because, in this way, we do not need a subsequent shape refinement. Furthermore, semantic (per-pixel) cell segmentation requires a subsequent pixel grouping that can result in segmentation errors such as falsely merging bordering cells. Since the images portray situations of very crowded cells,  such errors would be very likely to happen in our case. Star-convex polygons provide a much better shape representation that overcomes these difficulties.
\end{enumerate}

\begin{figure}[p]
    \centering
    \includegraphics[width=\textwidth]{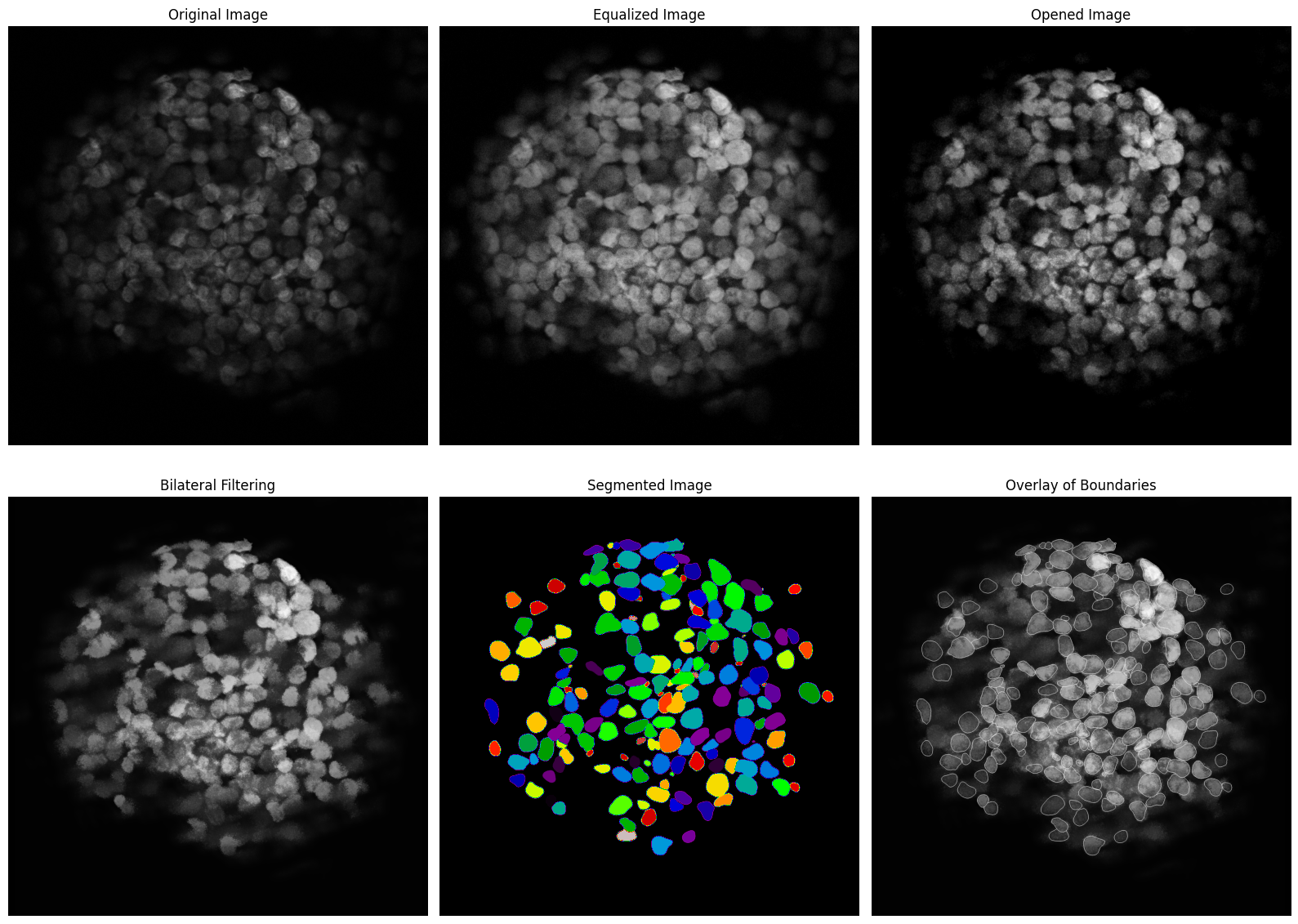}
    \caption{\emph{Image processing of the nuclei fluorescence channel (DAPI).} From left to right and top to bottom, the different steps in the processing of an image are followed. The borders of the identified structures over the original image are shown in the bottom right corner. The darker, unrecognized nuclei correspond to cells that are outside, in this case behind, the focal plane.}
    \label{fig: nuclei processing}
\end{figure}

The complete process on DAPI channel is depicted in Fig. \ref{fig: nuclei processing}. The original image is in the upper left corner. The lower right image is an enhanced representation highlighting the reconstructed cell boundaries.

\subsubsection*{Tessellation for cell region identification}
\begin{figure}[!ht]
   \includegraphics[width=\textwidth]{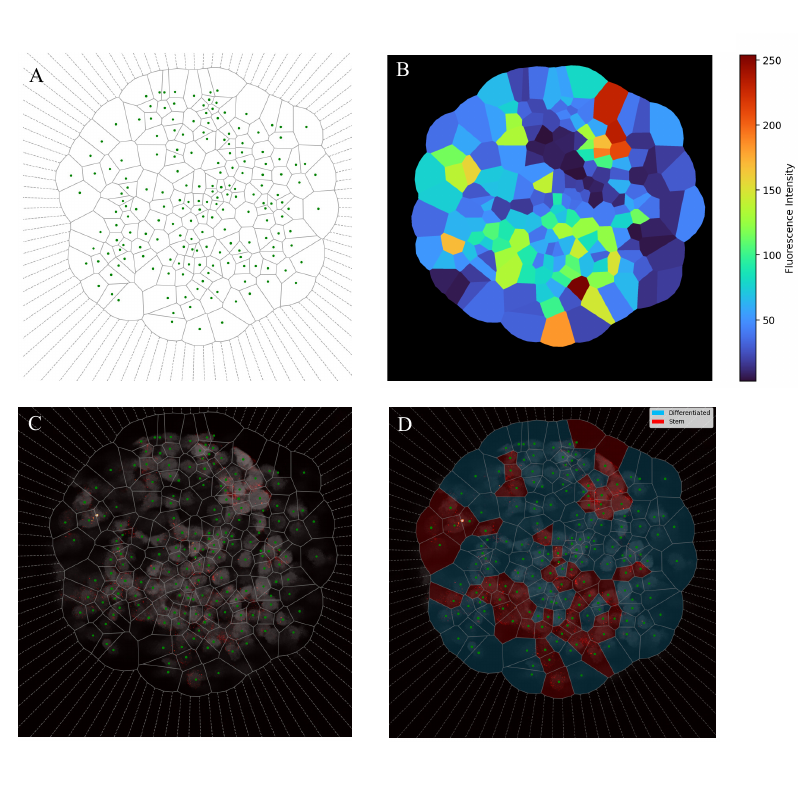}
   \caption{\emph{From tessellation to CSC recognition.} \textbf{A} Voronoi tessellation. The centroids of the cells, green dots, were used to estimate the cell area. \textbf{B} Regions colored according to SOX2 fluorescence intensity. For each region of the tessellation, the sum of its SOX2 fluorescence intensity is computed. \textbf{C} Tessellation overlaid with nuclei (gray scale) and SOX2 (red palette) channels. The boundaries between Voronoi regions are plotted with gray lines. \textbf{D} Tessellation regions are colored in red for CSCs and blue for DCCs, according to their SOX2 fluorescence content. If the fluorescence surpasses the threshold given by the GMM, the cell is considered a stem one.}
   \label{fig: tessellation to clustered}
\end{figure}

Since the SOX2 marker is not necessarily bound to the nucleus of the cell, but rather to its cytoplasm, we need a way of assigning each point in space (i.e., each pixel) to a single cell to further associate the fluorescence of the marker with a given cell. To do this, we approximated the division of the space associated with each cell by a Voronoi tessellation. We used the centers of the segmented objects filtering out the ones corresponding to cells that do not belong to the spheroid when needed. The result is shown in Fig. \ref{fig: tessellation to clustered}A. It is worth mentioning that we used artificial points, uniformly distributed in a circumference enclosing the spheroid, to limit the area of the cells in its border.

We thus associated to each cell, the sum of the fluorescence intensity of each pixel of the SOX2 channel contained inside its corresponding Voronoi region. The result of this association is shown in Fig. \ref{fig: tessellation to clustered}B as a color scale depicting the amount of SOX2 in each Voronoi region. To better visualize the tessellation, we overlayed it on the nuclei fluorescence and the cleaned SOX2 channel in Fig. \ref{fig: tessellation to clustered}C.

\subsubsection*{Stemness threshold in SOX2 fluorescence intensity}

To identify the CSCs we need to define a threshold in the intensity of SOX2 fluorescence that divides non-stem from stem cells. We implemented a clustering algorithm to group the intensities of the cells and an \emph{elbow plot} using k-means clustering. Its result confirmed the suitability of our model with two cell phenotypes. Thus, we considered a fixed number of two groups where data points were assigned according to a Gaussian Mixture Model (GMM) fitted to the data. In short, we assumed that the points were generated by two normal distributions, fitting their means and standard deviations using the maximum likelihood criterion. Additional details can be found in the  \href{https://scikit-learn.org/stable/modules/generated/sklearn.mixture.GaussianMixture.html}{documentation}.

\begin{figure}
    \centering
    \includegraphics[width=1.0\textwidth]{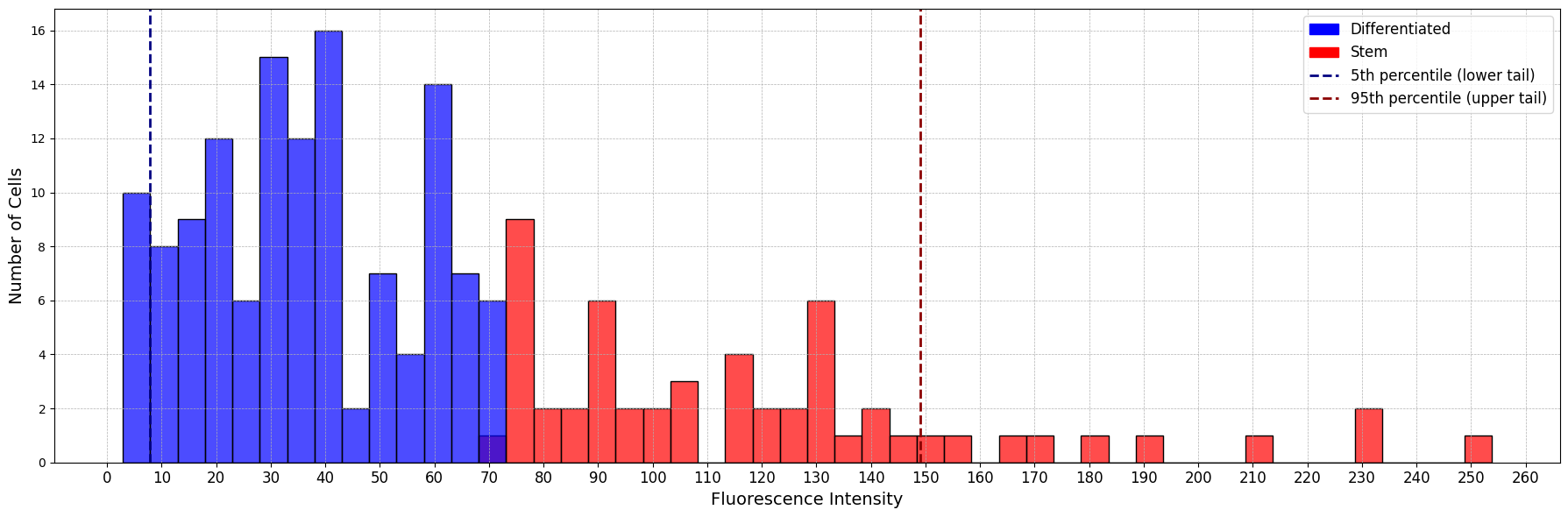}
    \caption{\emph{Histogram of SOX2 fluorescence intensities, colored by the GMM clustering.} Intensity values were clustered into differentiated and stem cells by fitting a GMM. Regions with intensities falling in the 5th  and  95th percentile were automatically considered as differentiated and stem respectively and were not taken into account when fitting the GMM.}
    \label{fig: filtered GMM histogram}
\end{figure}

Since the fitting of the distributions is sensitive to extreme values, the outliers, below 5th and above 95th percentiles, were directly assigned to their corresponding category, differentiated for the lowest and stem for the highest values, and were not taken into account when fitting the GMM. An example of the histogram of SOX2 fluorescence intensities for \textsf{Sph1,slice 3} is shown in Fig. \ref{fig: filtered GMM histogram}. The threshold value $V=71.4$ implies that from a total of $N=183$ cells present in the spheroid, $S_V=55$ are stem cells, a fraction of $f_V\simeq0.3$ of the total.
For every case, we checked that the random seed of the clustering algorithm had not modified the threshold significantly. Usually, there are a couple of values to which the threshold converges. Checking for robustness of the clustering means that these values are close to each other, regardless of the random state used and the clustering algorithm. If these values vary a lot or are not close to each other, the clustering becomes unreliable. For the case of our example, both GMM and K-means yield a threshold of 71.4, for a large number of random seeds.

A more intuitive way of visualizing this result is taking Fig. \ref{fig: tessellation to clustered}C and coloring each region according to its clustering. This is shown in Fig. \ref{fig: tessellation to clustered}D where superimposed to the confocal image, the regions corresponding to stem cells are colored in red, and the regions corresponding to differentiated cells are colored in blue. 

The final result of the image processing is shown in Fig. \ref{fig: blue red}. There just the Voronoi regions are depicted as being colored according to their category. The image depicted in Fig. \ref{fig: blue red}C corresponds to the one in Fig. \ref{fig: tessellation to clustered}D.

\subsubsection*{CSC connectedness and path forming} \label{sss: csc connectedness}
The location of the CSCs seems to be far from random by visual inspection of Fig. \ref{fig: blue red}, since the paths that they form, break the homogeneity of their distribution. This intuition is confirmed by the graph of a \emph{Delaunay triangulation}, the dual lattice of the Voronoi tessellation. This is nothing else than the graph given by using the centroids of the Voronoi as nodes, and adding a link between two nodes if the corresponding cells in the Voronoi diagram are neighboring cells. For instance, in Fig. \ref{fig: Delaunay sph1 slice3}A we depict the network corresponding to \textsf{ Sph1, slice 3}, the red and blue dots correspond to CSCs and DCCs respectively. Also we studied the subgraph with just the CSCs shown in Fig. \ref{fig: Delaunay sph1 slice3}B. This graph is obtained by removing all the blue nodes and all the links connected to them. We used these isolated paths, to assess if two CSCs are neighboring by chance. Particularly, we look if the nodes of the network have preferential attachment for nodes of the same type. We calculated the \emph{homophily} of the graph, i.e. how much more likely are CSCs to be connected between them, than to DCCs. This was achieved by measuring, on the graphs, the following parameters:

\begin{itemize}
    \item \emph{Assortativity coefficient}: The measurement of preferential attachment that is used in network science. It is calculated as the Pearson correlation coefficient of the attribute indicating the type of cell, between pairs of linked nodes. The coefficient equals 1 for perfect assortativity (perfect preferential attachment), 0 for non-assortative networks (i.e., no preferential attachment), and -1 for perfectly disassortative networks (perfect heterophily).
    \item \emph{Homophily ratio}: Is the fraction of the edges of the graph that link nodes of the same kind. The calculation is direct, dividing the number of edges connecting cells of the same type by the total number of edges. 
    \item \emph{Number of stem-connected components}: At the subgraph formed by considering only the stem cells of the original graph and the links between them (stem subgraph), is the number of separated clusters. If stem cells form clusters, the number of connected components will be smaller than expected for a random distribution.
    \item \emph{Degree distribution of the stem subgraph}: In the stem subgraph, it is the distribution of the number of stem cells that are connected to each stem cell. For instance, if the mean degree is higher than expected for randomly distributed stem cells, it means that stem cells are more connected between them than what would be expected from the random case.
\end{itemize}
\lstinline{NetworkX}'s\cite{SciPyProceedings_11} implementation was used for measuring all parameters, except for the homophily ratio. For more details consult the package's \href{https://networkx.org/documentation/stable/reference/algorithms/generated/networkx.algorithms.assortativity.attribute_assortativity_coefficient.html#networkx.algorithms.assortativity.attribute_assortativity_coefficient}{documentation}). 

We also compared with the case in which CSCs were placed at random positions. On the same networks obtained by Delaunay triangulation on the pictures, and leaving the connections untouched, we randomly redistributed the CSC in the graph. An example is shown in Fig. \ref{fig: Delaunay sph1 slice3}C where the red and blue dots are now randomly placed in other nodes, as seen when compared with panel A. In panel D we depicted the stem subgraph corresponding to the randomized network of panel C. Note that the number of connected components has increased from 5 in panel B to 13 for this case, a situation that occurs for most randomized networks. We performed this relocation of the whole set of CSCs 10000 times and measured the mentioned parameters joined with their corresponding standard deviations. The results comparing both, experimental and randomized networks, are reported in Table \ref{tab: ensemble statistics}. We also performed statistical tests (z-test via \textsf{statsmodels} implementation \cite{seabold2010statsmodels}) to decide whether a given value could be obtained from the distribution of the sample, i.e. if the parameter is significantly different from the random case. In every case, we used a significance $\alpha_{t}=0.001$, but results are robust against even lower values of $\alpha_{t}$.

\bibliography{sample}

\begin{thebibliography}{10}
\urlstyle{rm}
\expandafter\ifx\csname url\endcsname\relax
  \def\url#1{\texttt{#1}}\fi
\expandafter\ifx\csname urlprefix\endcsname\relax\def\urlprefix{URL }\fi
\expandafter\ifx\csname doiprefix\endcsname\relax\def\doiprefix{DOI: }\fi
\providecommand{\bibinfo}[2]{#2}
\providecommand{\eprint}[2][]{\url{#2}}

\bibitem{Al-Hajj}
\bibinfo{author}{Al-Hajj, M.}, \bibinfo{author}{Wicha, M.~S.},
  \bibinfo{author}{Benito-Hernandez, A.}, \bibinfo{author}{Morrison, S.~J.} \&
  \bibinfo{author}{Clarke, M.~F.}
\newblock \bibinfo{journal}{\bibinfo{title}{Prospective identification of
  tumorigenic breast cancer cells}}.
\newblock {\emph{\JournalTitle{Proc.Nat.Ac.Sc.}}}
  \textbf{\bibinfo{volume}{100}}, \bibinfo{pages}{3983--3988}
  (\bibinfo{year}{2003}).

\bibitem{Kakarala}
\bibinfo{author}{Kakarala, M.} \& \bibinfo{author}{Wicha, M.}
\newblock \bibinfo{journal}{\bibinfo{title}{Implications of the cancer
  stem-cell hypothesis for breast cancer prevention and therapy}}.
\newblock {\emph{\JournalTitle{J. Clin. Oncol.}}}
  \textbf{\bibinfo{volume}{26}}, \bibinfo{pages}{2813--2820},
  \doiprefix\url{10.1200/JCO.2008.16.3931} (\bibinfo{year}{2008}).

\bibitem{Dontu}
\bibinfo{author}{Dontu, G.} \emph{et~al.}
\newblock \bibinfo{journal}{\bibinfo{title}{In vitro propagation and
  transcriptional profiling of human mammary stem/progenitor cells}}.
\newblock {\emph{\JournalTitle{Genes Dev.}}} \textbf{\bibinfo{volume}{17}},
  \bibinfo{pages}{1253--1270}, \doiprefix\url{10.1101/gad.1061803}
  (\bibinfo{year}{2003}).

\bibitem{Ehmsen}
\bibinfo{author}{Ehmsen, S.} \emph{et~al.}
\newblock \bibinfo{journal}{\bibinfo{title}{Increased cholesterol biosynthesis
  is a key characteristic of breast cancer stem cells influencing patient
  outcome}}.
\newblock {\emph{\JournalTitle{Cell Rep.}}} \textbf{\bibinfo{volume}{27}},
  \bibinfo{pages}{3927--3938} (\bibinfo{year}{2019}).

\bibitem{Pastrana}
\bibinfo{author}{Pastrana~E, D.~F., Silva-Vargas~V}.
\newblock \bibinfo{journal}{\bibinfo{title}{Eyes wide open: a critical review
  of sphere-formation as an assay for stem cells}}.
\newblock {\emph{\JournalTitle{Cell Stem Cell}}} \textbf{\bibinfo{volume}{8}},
  \bibinfo{pages}{486--498}, \doiprefix\url{10.1016/j.stem.2011.04.007.}
  (\bibinfo{year}{2011}).

\bibitem{Kessel}
\bibinfo{author}{Kessel~SL, C.~L.}
\newblock \bibinfo{journal}{\bibinfo{title}{A high-throughput image cytometry
  method for the formation, morphometric, and viability analysis of
  drug-treated mammospheres}}.
\newblock {\emph{\JournalTitle{SLAS Discov.}}} \textbf{\bibinfo{volume}{25}},
  \bibinfo{pages}{723--733}, \doiprefix\url{10.1177/2472555220922817}
  (\bibinfo{year}{2020}).

\bibitem{Brady2019}
\bibinfo{author}{Brady, R.} \& \bibinfo{author}{Enderling, H.}
\newblock \bibinfo{journal}{\bibinfo{title}{Mathematical models of cancer: When
  to predict novel therapies, and when not to}}.
\newblock {\emph{\JournalTitle{Bull Math Biol}}} \bibinfo{pages}{3722--3731},
  \doiprefix\url{10.1007/s11538-019-00640-x} (\bibinfo{year}{2019}).

\bibitem{barberis2021percolation}
\bibinfo{author}{Barberis, L.}
\newblock \bibinfo{journal}{\bibinfo{title}{Radial percolation reveals that
  {{Cancer Stem Cells}} are trapped in the core of colonies}}.
\newblock {\emph{\JournalTitle{Papers in Physics}}}
  \textbf{\bibinfo{volume}{13}}, \bibinfo{pages}{130002--130002},
  \doiprefix\url{10.4279/pip.130002} (\bibinfo{year}{2021}).

\bibitem{Zanoni2020}
\bibinfo{author}{Zanoni, M.} \emph{et~al.}
\newblock \bibinfo{journal}{\bibinfo{title}{Modeling neoplastic disease with
  spheroids and organoids}}.
\newblock {\emph{\JournalTitle{J Hematol Oncol}}}
  \doiprefix\url{10.1186/s13045-020-00931-0} (\bibinfo{year}{2020}).

\bibitem{Rolver2019}
\bibinfo{author}{Rolver, M.~G.}, \bibinfo{author}{Elingaard-Larsen, L.~O.} \&
  \bibinfo{author}{Pedersen, S.~F.}
\newblock \bibinfo{journal}{\bibinfo{title}{Assessing cell viability and death
  in 3d spheroid cultures of cancer cells}}.
\newblock {\emph{\JournalTitle{J Vis Exp}}} \doiprefix\url{10.3791/59714}
  (\bibinfo{year}{2019}).

\bibitem{benitez2019}
\bibinfo{author}{Ben{\'i}tez, L.}, \bibinfo{author}{Barberis, L.} \&
  \bibinfo{author}{Condat, C.~A.}
\newblock \bibinfo{journal}{\bibinfo{title}{Modeling tumorspheres reveals
  cancer stem cell niche building and plasticity}}.
\newblock {\emph{\JournalTitle{Physica A}}} \textbf{\bibinfo{volume}{533}},
  \bibinfo{pages}{121906}, \doiprefix\url{10.1016/j.physa.2019.121906}
  (\bibinfo{year}{2019}).

\bibitem{barberis2021diff}
\bibinfo{author}{Barberis, L.~M.}, \bibinfo{author}{Benitez, L.} \&
  \bibinfo{author}{Condat, C.}
\newblock \bibinfo{journal}{\bibinfo{title}{Elucidating the {{Role Played}} by
  {{Cancer Stem Cells}} in {{Cancer Growth}}}}.
\newblock {\emph{\JournalTitle{MMSB}}} \textbf{\bibinfo{volume}{1}},
  \bibinfo{pages}{48--54} (\bibinfo{year}{2021}).

\bibitem{benitez2021}
\bibinfo{author}{Ben{\'i}tez, L.}, \bibinfo{author}{Barberis, L.},
  \bibinfo{author}{Vell{\'o}n, L.} \& \bibinfo{author}{Condat, C.~A.}
\newblock \bibinfo{journal}{\bibinfo{title}{Understanding the influence of
  substrate when growing tumorspheres}}.
\newblock {\emph{\JournalTitle{BMC Cancer}}} \textbf{\bibinfo{volume}{21}},
  \bibinfo{pages}{276}, \doiprefix\url{10.1186/s12885-021-07918-1}
  (\bibinfo{year}{2021}).

\bibitem{fotinos23}
\bibinfo{author}{Fotin{\'o}s, J.}, \bibinfo{author}{Barberis, L.} \&
  \bibinfo{author}{Condat, C.}
\newblock \bibinfo{journal}{\bibinfo{title}{Effects of a differentiating
  therapy on cancer-stem-cell-driven tumors}}.
\newblock {\emph{\JournalTitle{Journal of Theoretical Biology}}}
  \textbf{\bibinfo{volume}{572}}, \bibinfo{pages}{111563},
  \doiprefix\url{10.1016/j.jtbi.2023.111563} (\bibinfo{year}{2023}).

\bibitem{Condat2006}
\bibinfo{author}{Condat, C.~A.} \& \bibinfo{author}{Menchon, S.~A.}
\newblock \bibinfo{journal}{\bibinfo{title}{Ontogenetic growth of multicellular
  tumor spheroids}}.
\newblock {\emph{\JournalTitle{Physica A}}} \textbf{\bibinfo{volume}{371}},
  \bibinfo{pages}{76--79}, \doiprefix\url{10.1016/j.physa.2006.04.082}
  (\bibinfo{year}{2006}).

\bibitem{Delsanto2008}
\bibinfo{author}{Delsanto, P.~P.}, \bibinfo{author}{Condat, C.~A.},
  \bibinfo{author}{Pugno, N.}, \bibinfo{author}{Gliozzi, A.~S.} \&
  \bibinfo{author}{Griffa, M.}
\newblock \bibinfo{journal}{\bibinfo{title}{A multilevel approach to cancer
  growth modeling.}}
\newblock {\emph{\JournalTitle{Journal of Theoretical Biology}}}
  \textbf{\bibinfo{volume}{250}}, \bibinfo{pages}{16--24},
  \doiprefix\url{10.1016/j.jtbi.2007.09.023} (\bibinfo{year}{2008}).

\bibitem{Menchon2011}
\bibinfo{author}{Menchon, S.~A.} \& \bibinfo{author}{Condat, C.~A.}
\newblock \bibinfo{journal}{\bibinfo{title}{Quiescent cells: A natural way to
  resist chemotherapy}}.
\newblock {\emph{\JournalTitle{Physica A}}} \textbf{\bibinfo{volume}{390}},
  \bibinfo{pages}{3354–3361.}, \doiprefix\url{10.1016/j.physa.2011.05.009}
  (\bibinfo{year}{2011}).

\bibitem{Barberis2015}
\bibinfo{author}{Barberis, L.}, \bibinfo{author}{Pasquale, M.~A.} \&
  \bibinfo{author}{Condat, C.~A.}
\newblock \bibinfo{journal}{\bibinfo{title}{Joint fitting reveals hidden
  interactions in tumor growth.}}
\newblock {\emph{\JournalTitle{J. Theor. Biol.}}}
  \textbf{\bibinfo{volume}{365C}}, \bibinfo{pages}{420--432},
  \doiprefix\url{10.1016/j.jtbi.2014.10.038} (\bibinfo{year}{2015}).

\bibitem{chen2016}
\bibinfo{author}{Chen, Y.~C.} \emph{et~al.}
\newblock \bibinfo{journal}{\bibinfo{title}{High-throughput single-cell derived
  sphere formation for cancer stem-like cell identification and analysis}}.
\newblock {\emph{\JournalTitle{Scientific Reports}}}
  \textbf{\bibinfo{volume}{6}}, \bibinfo{pages}{1--12},
  \doiprefix\url{10.1038/srep27301} (\bibinfo{year}{2016}).

\bibitem{wang2016}
\bibinfo{author}{Wang, J.} \emph{et~al.}
\newblock \bibinfo{journal}{\bibinfo{title}{A novel method to limit breast
  cancer stem cells in states of quiescence, proliferation or differentiation:
  {{Use}} of gel stress in combination with stem cell growth factors}}.
\newblock {\emph{\JournalTitle{Oncology Letters}}}
  \textbf{\bibinfo{volume}{12}}, \bibinfo{pages}{1355--1360},
  \doiprefix\url{10.3892/ol.2016.4757} (\bibinfo{year}{2016}).

\bibitem{Leis}
\bibinfo{author}{Leis, O.} \emph{et~al.}
\newblock \bibinfo{journal}{\bibinfo{title}{Sox2 expression in breast tumours
  and activation in breast cancer stem cells}}.
\newblock {\emph{\JournalTitle{Oncogene}}} \textbf{\bibinfo{volume}{31}},
  \bibinfo{pages}{1354--1365} (\bibinfo{year}{2012}).

\bibitem{Marks2024}
\bibinfo{author}{Marks, M.~P.} \emph{et~al.}
\newblock \bibinfo{journal}{\bibinfo{title}{Role of
  hydroxymethylglutharyl-coenzyme a reductase in the induction of stem-like
  states in breast cancer}}.
\newblock {\emph{\JournalTitle{Journal of Cancer Research and Clinical
  Oncology}}}  (\bibinfo{year}{2024}).

\bibitem{Stines2013}
\bibinfo{author}{Stine, R.~R.} \& \bibinfo{author}{Matunis, E.~L.}
\newblock \bibinfo{journal}{\bibinfo{title}{Stem cell competition: finding
  balance in the niche}}.
\newblock {\emph{\JournalTitle{Trends Cell Biol}}} \bibinfo{pages}{357--364},
  \doiprefix\url{10.1016/j.tcb.2013.03.001} (\bibinfo{year}{2013}).

\bibitem{Johnston2009}
\bibinfo{author}{Johnston, L.~A.}
\newblock \bibinfo{journal}{\bibinfo{title}{Competitive interactions between
  cells: death, growth, and geography}}.
\newblock {\emph{\JournalTitle{Science}}} \bibinfo{pages}{1679--1682},
  \doiprefix\url{10.1126/science.1163862} (\bibinfo{year}{2009}).

\bibitem{Derks2023}
\bibinfo{author}{Derks, L. L.~M.} \& \bibinfo{author}{van Boxtel, R.}
\newblock \bibinfo{journal}{\bibinfo{title}{Stem cell mutations, associated
  cancer risk, and consequences for regenerative medicine}}.
\newblock {\emph{\JournalTitle{Cell Stem Cell}}} \bibinfo{pages}{1421--1433},
  \doiprefix\url{10.1016/j.stem.2023.09.008} (\bibinfo{year}{2023}).

\bibitem{Fisher2020}
\bibinfo{author}{Fisher, M.~F.} \& \bibinfo{author}{Rao, S.~S.}
\newblock \bibinfo{journal}{\bibinfo{title}{Three-dimensional culture models to
  study drug resistance in breast cancer}}.
\newblock {\emph{\JournalTitle{Biotechnol Bioeng}}}
  \bibinfo{pages}{2262--2278}, \doiprefix\url{10.1002/bit.27356}
  (\bibinfo{year}{2020}).

\bibitem{Rodriguez2018}
\bibinfo{author}{Rodríguez, C.~E.} \emph{et~al.}
\newblock \bibinfo{journal}{\bibinfo{title}{Breast cancer stem cells are
  involved in trastuzumab resistance through the her2 modulation in 3d
  culture}}.
\newblock {\emph{\JournalTitle{J Cell Biochem}}} \bibinfo{pages}{1381--1391},
  \doiprefix\url{10.1002/jcb.26298} (\bibinfo{year}{2018}).

\bibitem{Olive1994}
\bibinfo{author}{Olive, P.~L.} \& \bibinfo{author}{Durand, R.~E.}
\newblock \bibinfo{journal}{\bibinfo{title}{Drug and radiation resistance in
  spheroids: cell contact and kinetics}}.
\newblock {\emph{\JournalTitle{Cancer Metastasis Rev}}}
  \bibinfo{pages}{121--138}, \doiprefix\url{10.1007/BF00689632}
  (\bibinfo{year}{1994}).

\bibitem{scikit-image}
\bibinfo{author}{van~der Walt, S.} \emph{et~al.}
\newblock \bibinfo{journal}{\bibinfo{title}{scikit-image: image processing in
  {P}ython}}.
\newblock {\emph{\JournalTitle{PeerJ}}} \textbf{\bibinfo{volume}{2}},
  \bibinfo{pages}{e453}, \doiprefix\url{10.7717/peerj.453}
  (\bibinfo{year}{2014}).

\bibitem{schmidt2018}
\bibinfo{author}{Schmidt, U.}, \bibinfo{author}{Weigert, M.},
  \bibinfo{author}{Broaddus, C.} \& \bibinfo{author}{Myers, G.}
\newblock \bibinfo{title}{Cell detection with star-convex polygons}.
\newblock In \emph{\bibinfo{booktitle}{Medical Image Computing and Computer
  Assisted Intervention--MICCAI 2018: 21st International Conference, Granada,
  Spain, September 16-20, 2018, Proceedings, Part II 11}},
  \bibinfo{pages}{265--273} (\bibinfo{organization}{Springer},
  \bibinfo{year}{2018}).

\bibitem{SciPyProceedings_11}
\bibinfo{author}{Hagberg, A.~A.}, \bibinfo{author}{Schult, D.~A.} \&
  \bibinfo{author}{Swart, P.~J.}
\newblock \bibinfo{title}{Exploring network structure, dynamics, and function
  using networkx}.
\newblock In \bibinfo{editor}{Varoquaux, G.}, \bibinfo{editor}{Vaught, T.} \&
  \bibinfo{editor}{Millman, J.} (eds.) \emph{\bibinfo{booktitle}{Proceedings of
  the 7th Python in Science Conference}}, \bibinfo{pages}{11 -- 15}
  (\bibinfo{address}{Pasadena, CA USA}, \bibinfo{year}{2008}).

\bibitem{seabold2010statsmodels}
\bibinfo{author}{Seabold, S.} \& \bibinfo{author}{Perktold, J.}
\newblock \bibinfo{title}{statsmodels: Econometric and statistical modeling
  with python}.
\newblock In \emph{\bibinfo{booktitle}{9th Python in Science Conference}}
  (\bibinfo{year}{2010}).

\end{thebibliography}

\section*{Acknowledgements }
The authors thank Dr. C. A. Condat for insightful discussions and comments. The confocal imaging was performed by Pablo Pomata at IBYME's Microscopy Service.

\section*{Funding}The theoretical work was supported by SECyT-UNC (Project 113/17) and CONICET (PIP 11220110100794), Argentina. Experiments were supported by  Consejo Nacional de Investigaciones Científicas y Técnicas (CONICET), Argentina, Fondo para la Investigación Científica y Tecnológica (FONCYT), Argentina, René Bigand Foundation, Argentina, René Barón Foundation, Argentina and Williams Foundation, Argentina.

\section*{Author contributions statement}

JF developed and implemented the image processing pipeline, LB and LV conceived the general idea. These authors analyzed the results and wrote the manuscript. MP carried out the experiment.

\section*{Additional information}
\textbf{Competing interests}: The authors declare no competing interests.

\noindent \textbf{Accession codes}: All code and data are accessible through the GitHub repository: \href{https://github.com/JeroFotinos/experimental_image_analysis}{\includegraphics[width=1em]{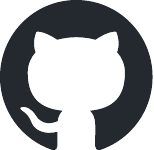} \nolinkurl{experimental_image_analysis}}



\end{document}